\documentclass[aps,showpacs,preprintnumbers,amsmath,amssymb]{revtex4}
\usepackage{dcolumn}
\usepackage{multirow}
\usepackage[dvips]{epsfig}
\def \be {\begin{equation}}
\def \ee {\end{equation}}
\def \bea {\begin{eqnarray}}
\def \eea {\end{eqnarray}}

\begin{document}
\baselineskip=0.8 cm
\title{\bf Chaotic motion of scalar particle coupling to Chern-Simons invariant in Kerr black hole spacetime}
\author{Xuan Zhou $^{1}$, Songbai Chen$^{1,2}$\footnote{Corresponding author: csb3752@hunnu.edu.cn}, Jiliang Jing$^{1,2}$ \footnote{jljing@hunnu.edu.cn}}

\affiliation{ $ ^1$ Department of Physics, Key Laboratory of Low Dimensional Quantum Structures
and Quantum Control of Ministry of Education, Synergetic Innovation Center for Quantum Effects and Applications, Hunan
Normal University,  Changsha, Hunan 410081, People's Republic of China
\\
$ ^2$Center for Gravitation and Cosmology, College of Physical Science and Technology, Yangzhou University, Yangzhou 225009, People's Republic of China}

\begin{abstract}
\baselineskip=0.6 cm
\begin{center}
{\bf Abstract}
\end{center}

We present firstly  the equation of motion for the test scalar particle coupling to the Chern-Simons invariant in Kerr black hole spacetime by the short-wave approximation.  We have analyzed the dynamical behaviors of the test coupled particles
by applying techniques including Poincar\'e sections, fast Lyapunov exponent indicator, bifurcation diagram and basins of attraction.  It is shown that there exists chaotic phenomenon in the motion of scalar particle interacted
with the Chern-Simons invariant in a Kerr black hole spacetime. With the increase of the coupling strength, the motion of the coupled particles for the chosen parameters first undergoes a series of transitions betweens chaotic motion and regular motion and then falls into horizon or escapes to spatial infinity. Thus, the coupling between scalar particle and Chern-Simons invariant yields the richer dynamical behavior of scalar particle in a Kerr black hole spacetime.

\end{abstract}

\maketitle
\newpage

\section{Introduction}

Chaos is a kind of very complex motions with high sensitivity to initial conditions, which occurs in a definiteness system with nonlinear interactions. One of the most important feature of chaos is that the tiny errors in the chaotic motion
grow at an exponential rate, which leads to that the motion differs totally from what it would be
without these errors \cite{Sprott, Ott,Brown,Brown1}. Thus, it is very difficult to make a long-term prediction to the motion of a chaotic dynamical system. This implies that chaotic systems own many novel properties not shared by the linear dynamical systems due to the nonlinear interactions, which triggers a lot of attention to study  chaotic dynamics in various
physical fields.

In general relativity, the chaotic behavior does not emerge in the geodesic motion of particle in the usual Kerr-Newman-like
black hole spacetimes \cite{Carter} because the geodesic equation of particle is variable-separable and the dynamical
system is integrable. In order to ensure that the dynamical system of particle is non-integrable
and study further its chaotic behavior, we must resort to some spacetimes with complex geometrical structures or
introduce some extra interactions. In this way, the chaotic orbits of particle have been investigated
in the perturbed Schwarzschild spacetime \cite{Bombelli,Aguirr}, or
in multi-black hole spacetimes \cite{Cornish,Hanan} , or in the spacetime of a black hole immersed in
a magnetic field \cite{Karas,Ldan}, or in the accelerating and rotating black hole spacetime
\cite{schen1}, or in the background of a non-standard rotating black hole \cite{Gair,Contopoulos0,Lukes0,Dubeibe0,Gueron0}. Moreover, chaotic behaviors of ring strings have been found in Schwarzschild black hole spacetime \cite{Frolov}, and in the AdS black hole spacetimes \cite{Zayas, ZzMa}. By introducing the extra interaction with Einstein tensor, the chaotic dynamics of a test scalar particle has also been studied in Schwarzschild-Melvin black hole spacetime \cite{mschen2}. Moreover, the chaotic  phenomenon also appear in the string loop dynamics around black hole \cite{kolo}, the charged particle motion in the
magnetosphere \cite{kolo1}, and in the particle motion very near to the horizon \cite{kolo2,kolo3}.

Einstein's general relativity is considered probably the most beautiful of all existing
physical theories, which has successfully passed a series of observational and experimental
tests \cite{mwill}. However, it is believed widely that Einstein's theory  may not be the final theory to describe
the gravitational field due to its incompatibility with quantum theory and
motivations from cosmology. Thus, there has been interest in
the study of possible extensions to Einstein's theory of gravity.
One of the most promising extensions of
general relativity is the Chern-Simons-modified gravity \cite{cs1,cs2,cs3}, in which the Einstein-Hilbert
action is modified by adding an extra interaction between scalar field and Chern-Simons invariant, which captures the leading-order
gravitational parity-violation. The interaction with Chern-Simons invariant is necessary in heterotic string theory as an anomaly canceling term to conserve unitarity \cite{cs4,cs5,cs6,cs7}. Moreover, it is required to ensure gauge invariance of the
Ashtekar variables in loop quantum gravity \cite{cs8}.

The Chern-Simons-modified gravity was firstly investigated in the non-dynamical formulation, where the Chern-Simons scalar field is not a dynamical field, but only an \textit{a priori} prescribed
function. In the non-dynamical Chern-Simons theory, it is found that a valid solution of spacetime exists only if the  Pontryagin density vanishes because the Cotton-York tensor would have a divergence in the nonzero Pontryagin density case \cite{dcs1,dcs2,dcs3}. Thus, the non-dynamical Chern-Simons theory is quite artificial and is always treated only as a toy model used to obtain some insight in parity-violating theories of gravity. Based on this consideration, a lot of attention has been focused on the dynamical Chern-Simons modified gravity where the scalar field is assumed to own its kinetic term and dynamical evolution equation \cite{dcs1,dcs4,dcs6}. In this dynamical theory, the scalar-field stress-energy and the Cotton-York tensors are separately conserved  as the Pontryagin density vanishes. However, for the nonzero Pontryagin density,
one can find that the divergence of the Cotton-York tensor is precisely balanced by the divergence of the scalar field stress-energy tensor \cite{dcs4,dcs6}, which differs from that in the non-dynamical Chern-Simons theory. It is of interest to study the properties of black holes in the dynamical Chern-Simons  modified gravity. In general, it is difficult to obtain an analytical black hole solution in this modified gravity because the Einstein's field equation and the dynamical equation of scalar field must be satisfied simultaneously.  Thus, the analytical solution of a rotating black hole in dynamical Chern-Simons gravity have been obtained only in the small-coupling and/or
slow-rotation limit \cite{dcs1,dcs6,Ali,Stein,Yagi,McNees}.  The observable effects of small Chern-Simons coupling parameter in such black hole spacetimes have been studied including black hole shadow \cite{Amarilla} and strong gravitational lensing \cite{Chensb1}. Without any perturbational expansion, a numerical solution of a rotating black hole is obtained in \cite{Delsate}, which is helpful to understand some important features emerging in the fast spinning and/or large coupling regimes in dynamical Chern-Simons modified gravity. Moreover, the stability of black holes under gravitational perturbations have also been studied in dynamical Chern-Simons modified gravity \cite{Card1,Card2,Card3}.

Most of above literature on dynamical Chern-Simons gravity focus mainly on the linear coupling case where the coupling term with Chern-Simons invariant is proportional to the scalar field.
Inspired by the phenomenon of spontaneous scalarization recently discussed in the quadratic scalar-Gauss-Bonnet
gravity \cite{Antoniou,Doneva,Silva,Antoniou2}, Gao \textit{et al} \cite{Yuan} generalized it to the dynamical Chern-Simons modified gravity and considered the case where the Chern-Simons
invariant is coupled to the quadratic function of the dynamical scalar field. And then they studied the scalar perturbation around a Kerr black hole and found that the black hole
becomes unstable under linear perturbations in a certain regions of the parameter space.
In this paper, we want to study the effects of such quadratic coupling on the motion of a test scalar particle in a Kerr black hole background.  In order to reach this purpose, we here assume the dynamical Chern-Simons scalar field as a perturbational field and then obtain the corrected Klein-Gordorn equation for the coupled scalar field. With the shortwave
approximation, we can get the equation of motion of the coupled scalar particle from the above corrected Klein-Gordorn equation. The quadratic coupling between the scalar field and the Chern-Simons invariant results in that the equation of motion for the coupled scalar particle is not variable-separable, which means that the chaotic behavior could appear in the motion of the scalar particle. It is natural to study how such coupling affects chaotic behaviors of scalar particles in black hole spacetimes.

The paper is organized as follows. In Sect. II, we obtain the geodesic equation of a test scalar particle coupled to the
Chern-Simons invariant in the Kerr black hole spacetime by the short-wave approximation. In Sect. III, we investigate the
chaotic phenomenon in the motion of the scalar particle coupled to the Chern-Simons invariant by techniques including the Poincar\'e section, the fast Lyapunov indicator,  and bifurcation diagram. We probe the effects of this coupling together with black hole spin parameter on the chaotic behavior of a coupled scalar particle. Finally, we end the paper with a
summary.

\section{Geodesics of scalar particle coupling to Chern-Simons invariant in
Kerr black hole spacetime}
In this section, we will derive the equations of motion for a test scalar particle coupling to Chern-Simons invariant in the Kerr black hole spacetime. The simplest action containing the quadratic-scalar-Chern-Simons invariant coupling term  can be expressed as \cite{Yuan}
\begin{equation}
S=\int d^{4} x \sqrt{-g}\left[\frac{R}{16\pi G}-\frac{1}{2} \nabla_{\mu} \Phi \nabla^{\mu} \Phi-\mu^2 \Phi^2+\alpha\;  ^{*}R R \Phi^2\right],\label{act1}
\end{equation}
where $R$ is Ricci curvature scalar and $G$ is Newton's constant. $\Phi$ is a scalar field with mass $\mu$. The Chern-Simons invariant $^{*} R R$ is a topological invariant, which is also called the Pontryagin density and can be expressed as
\begin{equation}
^{*} R R=\frac{1}{2} \epsilon^{\alpha \beta \gamma \delta} R_{\;\nu \gamma \delta}^{\mu} R_{\;\mu \alpha \beta}^{\nu},
\end{equation}
where $\epsilon^{\alpha \beta \gamma \delta}$ is the well-known Levi-Civita tensor and $R_{\;\nu \gamma \delta}^{\mu}$ is the usual Riemann curvature tensor. The parameter $\alpha$ is a coupling constant with dimension of length squared.
Varying the action (\ref{act1}) with respect to the metric $g_{\mu\nu}$, one can obtain the modified Einstein equation \cite{Yuan}
\begin{eqnarray}\label{einst0}
G^{\mu\nu}+64\alpha\pi G  C^{\mu\nu}=8\pi G T^{\mu\nu},
\end{eqnarray}
where $G^{\mu\nu}$ is the usual contravariant Einstein tensor,
\begin{eqnarray}\label{einst1}
 C^{\mu\nu}=\partial_{\sigma}\Phi^2\epsilon^{\sigma\beta\gamma(\mu}\partial_{\gamma}R^{\nu)}_{\;\;\;\beta}
 +\partial_{\beta}\partial_{\gamma}\Phi^2\;^{*}R^{\mu(\beta\gamma)\nu},\quad\quad\quad
 T^{\mu\nu}=\partial^{\mu}\Phi\partial^{\nu}\Phi-g^{\mu\nu}[\frac{1}{2}\partial_{\lambda}\Phi\partial^{\lambda}\Phi+\mu^2\Phi^2].
\end{eqnarray}
Similarly, varying the action (\ref{act1}) with respect to the scalar field  $\Phi$ , one can get the modified Klein-Gordon equation
\begin{equation}
\frac{1}{\sqrt{-g}}\frac{\partial}{\partial x^{\mu}}\bigg[\sqrt{-g}g^{\mu\nu}\frac{\partial \Phi}{\partial x^{\nu}}\bigg]+(2 \alpha^{*} R R-\mu^2)\Phi=0.\label{cKGordon}
\end{equation}
Here the curvature correction acts actually as an effective mass.
Assuming the amplitude $f$ of the scalar field is a small slowly-varying real and the derivative term $f_{;\mu}$ can be neglected, one can find that the tensors $C^{\mu\nu}$ and $T^{\mu\nu}$ in the  Einstein equation (\ref{einst0})  are second order small quantity $\mathcal{O}(f^2)$, but the Klein-Gordon equation (\ref{cKGordon}) is the first order equation of $f$. Thus, in the first order approximation of $f$, we find that the scalar field and its interaction with Chern-Simons invariant do not modify the background spacetime and then scalar field can be treated as a perturbational field. With these assumption and approximation, it is easy to find that the Kerr metric is a solution of the action (\ref{act1}), whose line element can be written in Boyer-Lindquist coordinates as
\begin{equation}
d s^{2}=-\frac{\Delta}{\rho^{2}}\left(d t-a \sin ^{2} \theta d \phi\right)^{2}+\frac{\sin ^{2} \theta}{\rho^{2}}\left[\left(r^{2}+a^{2}\right) d \phi-a d t\right]^{2}+\frac{\rho^{2}}{\Delta} d r^{2}+\rho^{2} d \theta^{2},
\end{equation}
with
\begin{equation}
\Delta \equiv r^{2}-2 M r+a^{2} ,~~~~~~~~~
\rho^{2} \equiv r^{2}+a^{2} \cos ^{2} \theta,
\end{equation}
where $M$, $a$ denote the mass and the spin parameter of black hole, respectively. The Chern-Simons invariant for Kerr black hole is
\begin{equation}\label{CSRR}
^* R R=\frac{96 a M^{2} r \cos \theta\left(3 r^{2}-a^{2} \cos ^{2} \theta\right)\left(r^{2}-3 a^{2} \cos ^{2} \theta\right)}{\left(r^{2}+a^{2} \cos ^{2} \theta\right)^{6}}.
\end{equation}
Since the Chern-Simons invariant $^* R R$ contains the linear term of $\cos\theta$,  it is not symmetric with respect to the equatorial plane.

It is well known that the Klein-Gordon equation (\ref{cKGordon}) is a wave equation.
In order to obtain the equation of motion of a test scalar particle from a wave equation (\ref{cKGordon}), we must adopt to the short-wave approximation because when the wavelength of the scalar particle is much smaller than the typical curvature, the particle aspect of scalar particle is dominated and its  wave aspect can be neglected. The similar treatment  are taken in \cite{shw1,shw2,shw3,shw4,shw5} to probe the propagation of photon under some interactions between  electromagnetic field and gravitational field.
In this approximation, the perturbational scalar field  $\Phi$ can be simplified as
\begin{equation}\label{shwave}
\Phi=fe^{iS},
\end{equation}
with a rapidly varying phase $S$. Since  the amplitude $f$ of the scalar field is small and slowly-varying, the derivative term $f_{;\mu}$ can be neglected because it is not dominated in this approximation.
The wave vector $\partial_{\mu}S$ can be regarded as the usual momentum $p_{\mu}$ of scalar particle. In this way, the corrected Klein-Gordon equation (\ref{cKGordon}) can be written as
\begin{equation}\label{sesan}
g^{\mu\nu}p_{\mu}p_{\nu}-2\alpha^{*} R R=-1.
\end{equation}
Here, we set the mass of the scalar particle $\mu=1$. The equation (\ref{sesan}) can be further rewritten as Hamilton-Jacobi equation \cite{shw6}
\begin{equation}\label{yelebi}
\frac{\partial S}{\partial \tau}+\mathcal{H}(x^{\mu},\frac{\partial S}{\partial x^{\mu}})=0,
\end{equation}
with $S=\frac{1}{2}\tau+x^{\mu}p_{\mu}$ and the Hamiltonian
\begin{equation}\label{Hamin}
\mathcal{H}(x^{\mu},\frac{\partial S}{\partial x^{\mu}})=\frac{1}{2}g^{\mu\nu}p_{\mu}p_{\nu}-\alpha ^{*} RR.
\end{equation}
Here $x^{\mu}$ is spacetime coordinate and $p_{\mu}$
is the corresponding canonical momentum. The quantity $\tau$ is  an affine parameter along the curve. The term $-\alpha ^{*} RR$ is an extra potential arising from the interaction between the scalar particle and  the Chern-Simons invariant, which disappears for a free test particle.

Since $g^{\mu\nu}$ and $^{*} R R$ in Hamiltonian (\ref{Hamin}) are only functions of the coordinates $r$ and $\theta$, the dynamical system of the scalar particle owns two cyclic coordinates $t$ and $\phi$. It means that there exist two conserved quantities, i.e., the energy $E$ and the $z$-component of the angular momentum $L$ of the timelike particle, which is similar to that in the non-coupling case. With these two conserved quantities and Hamilton canonical  equation
\begin{equation}\label{HaminEq}
\frac{dx^{\mu}}{d\tau}=\frac{\partial \mathcal{H}}{\partial p_{\mu}},\quad\quad\quad \frac{dp_{\mu}}{d\tau}=-\frac{\partial \mathcal{H}}{\partial x^{\mu}},
\end{equation}
one can obtain the geodesic equation for the scalar particle coupling to Chern-Simons invariant in Kerr black hole spacetime,
\begin{equation}\label{wfE1}
\dot{t}=\frac{{g}_{\phi \phi} E+{g}_{t \phi} L}{{g}_{t \phi}^{2}-{g}_{t t} {g}_{\phi \phi}},\quad\quad\quad
\dot{\varphi}=-\frac{{g}_{t \phi} E+{g}_{t t} L}{{g}_{t \phi}^{2}-{g}_{t t} {g}_{\phi \phi}},
\end{equation}
and
\begin{equation}\label{wfE2}
\ddot{r}=\frac{1}{2} g^{rr}\left(g_{t t, r} \dot{t}^{2}-g_{r r, r} \dot{r}^{2}+g_{\theta \theta, r} \dot{\theta}^{2}+g_{\phi \phi, r} \dot{\phi}^{2}+2 g_{t \phi, r} \dot{t} \dot{\phi}-2 g_{rr, \theta} \dot{r} \dot{\theta}+2 \alpha \frac{\partial ^{*} R R}{\partial r}\right),
\end{equation}
\begin{equation}\label{wfE3}
\ddot{\theta}=\frac{1}{2} g^{\theta \theta}\left(g_{t t, \theta} \dot{t}^{2}+g_{r r, \theta} \dot{r}^{2}-g_{\theta \theta, r} \dot{\theta}^{2}+g_{\phi \phi, \theta} \dot{\phi}^{2}+2 g_{t \phi,\theta} \dot{t} \dot{\phi}-2 g_{r \theta, r} \dot{r} \dot{\theta}+2 \alpha \frac{\partial^{*} R R}{\partial \theta}\right),
\end{equation}
with a constraint condition
\begin{equation}\label{Hcon}
h=g_{t t} \dot{t}^{2}+g_{r r} \dot{r}^{2}+g_{\theta \theta} \dot{\theta}^{2}+g_{\phi \phi} \dot{\phi}^{2}+2 g_{t \phi} \dot{t} \dot{\phi}+1-2 \alpha^{*} R R=0.
\end{equation}
As the Chern-Simons coupling vanishes, these equations reduce to those of the usual timelike particles in the Kerr black hole spacetime. In the non-zero coupling case, the complicated Chern-Simons invariant (\ref{CSRR}) results in that
the equation (\ref{Hcon}) is not variable-separable, which differs from that in the case without the Chern-Simons  coupling. Thus,  chaotic behavior could appear in motion of a scalar particle due to the interaction with  Chern-Simons invariant. In the next section, we will investigate the effect of the coupling parameter $\alpha$ on  motion of a coupled scalar particle in a Kerr black hole background.

\section{ Chaotic motion of scalar particles coupling to Chern-Simons invariant in
Kerr black hole spacetime}

\begin{figure}
\includegraphics[width=4cm ]{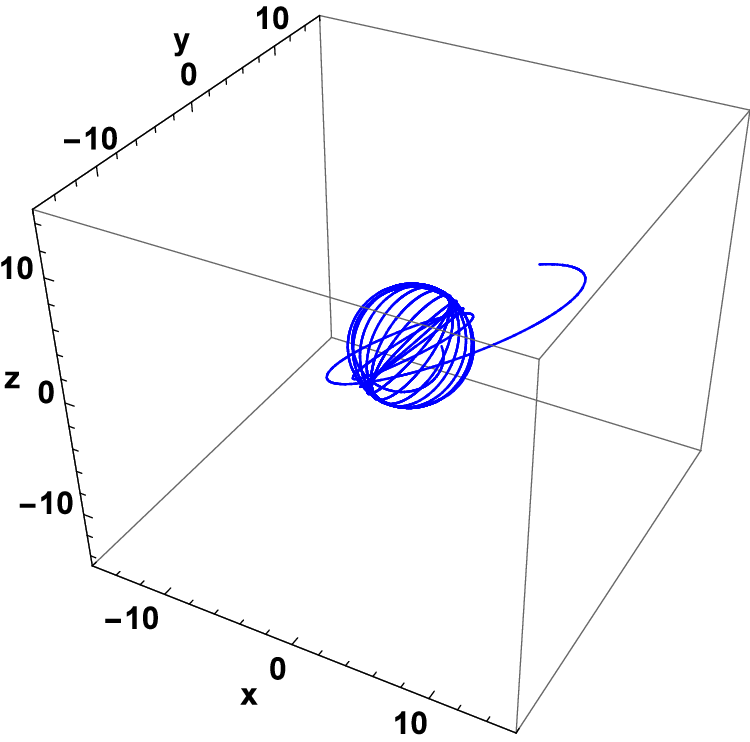}\includegraphics[width=4cm ]{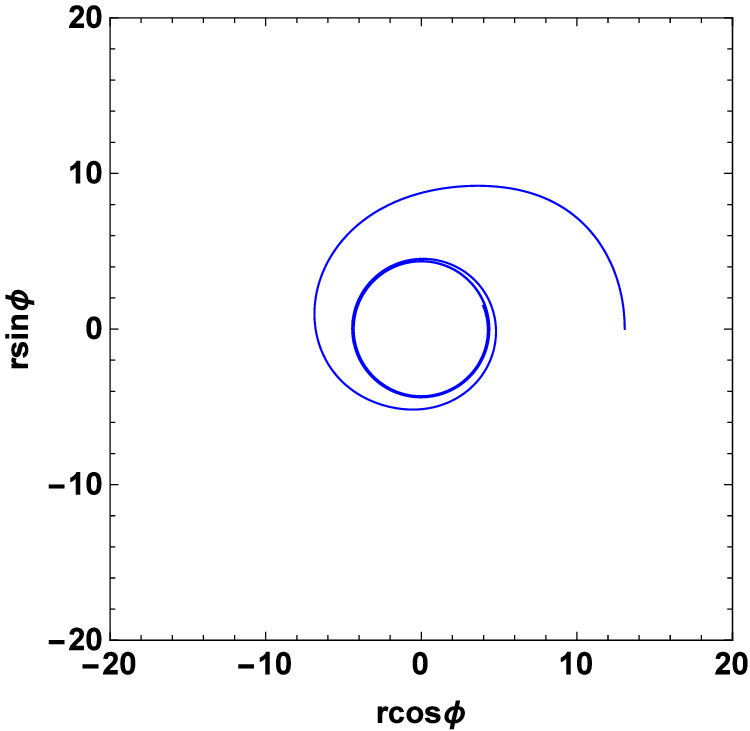}\includegraphics[width=4cm]{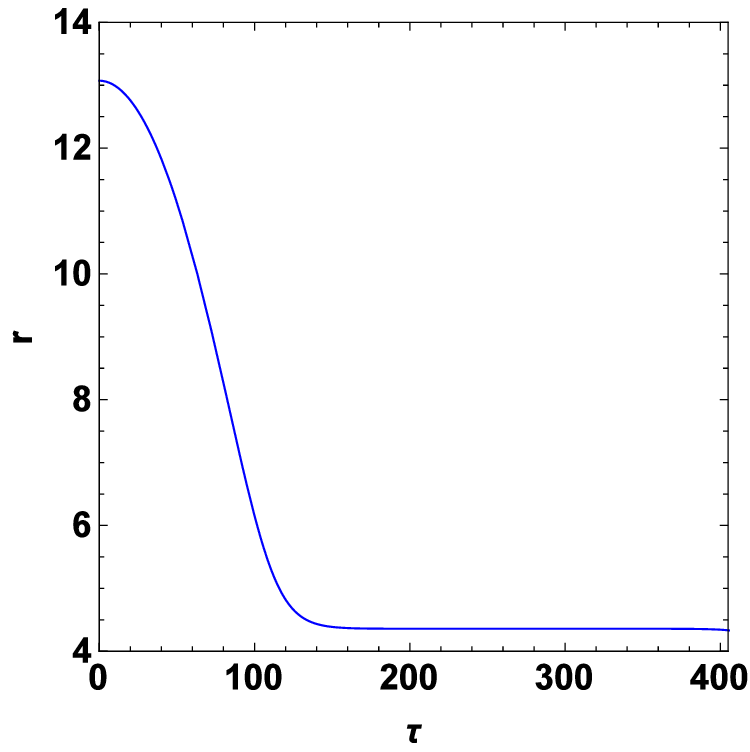}\includegraphics[width=4cm ]{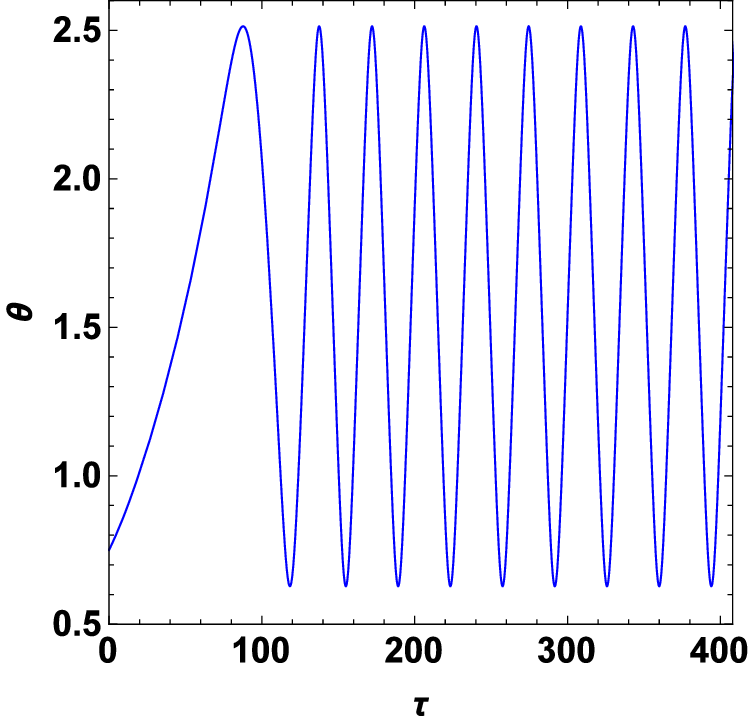}
\caption{The homoclinic orbit H3 in the Kerr spacetime in ref.\cite{Rana}. The system parameters are set to $a=0.2$, $E=0.95303$ and $L=2.05365$.}\label{fign1}
\end{figure}

It is well known that chaos is a class of very complex motion with high sensitivity to initial value. In general, detecting chaos in the dynamical system must resort to techniques including the Poincar\'e section, the fast Lyapunov indicator, and bifurcation diagram. However, it must be kept in mind that these techniques in themselves are not
rigorous proofs. There are some methods to prove the occurrence of chaos by exploring the transversal homoclinic orbits \cite{Rod,Rod1,Melnik,Melnik1,Bombelli}.
In this section, we firstly analyze homoclinic tangles  and then check the existence of chaos in the dynamical system (\ref{Hamin}). We rewrite the Hamiltonian (\ref{Hamin}) as two parts
\begin{eqnarray}\label{Haminpert}
 \mathcal{H}=\mathcal{H}_0+\mathcal{H}_1,
\end{eqnarray}
with
\begin{eqnarray}
\mathcal{H}_0&\equiv &\frac{1}{2}g^{\mu\nu}p_{\mu}p_{\nu}+\frac{1}{2}=\frac{1}{2}g^{rr}p_{r}p_{r}+\frac{1}{2}g^{\theta\theta}p_{\theta}p_{\theta}+\frac{1}{2\rho^2}
\bigg[\frac{(L-aE^2\sin^2\theta)^2}{\sin^{2}\theta}-\frac{[(r^2+a^2)E-aL]^2}{\Delta}\bigg]+\frac{1}{2},\nonumber\\
\mathcal{H}_1&\equiv &-\alpha ^{*} RR=-\frac{96 \alpha a M^{2} r \cos \theta\left(3 r^{2}-a^{2} \cos ^{2} \theta\right)\left(r^{2}-3 a^{2} \cos ^{2} \theta\right)}{\left(r^{2}+a^{2} \cos ^{2} \theta\right)^{6}}.
\end{eqnarray}
$\mathcal{H}_0$ is the Hamiltonian of the usual timelike geodesics particles in Kerr spacetime, $\mathcal{H}_1$ is the Hamiltonian arising from the interaction with Chern-Simons invariant. The homoclinic orbits in Kerr spacetime are analyzed in ref.\cite{Rana}. Here, without loss of generality, we take the homoclinic orbit H3 as an example to analyze homoclinic tangles under the interaction with Chern-Simons invariant. For convenience, the homoclinic orbit H3 in ref.\cite{Rana} is also shown in Fig.\ref{fign1}.
\begin{figure}
\includegraphics[width=5.5cm ]{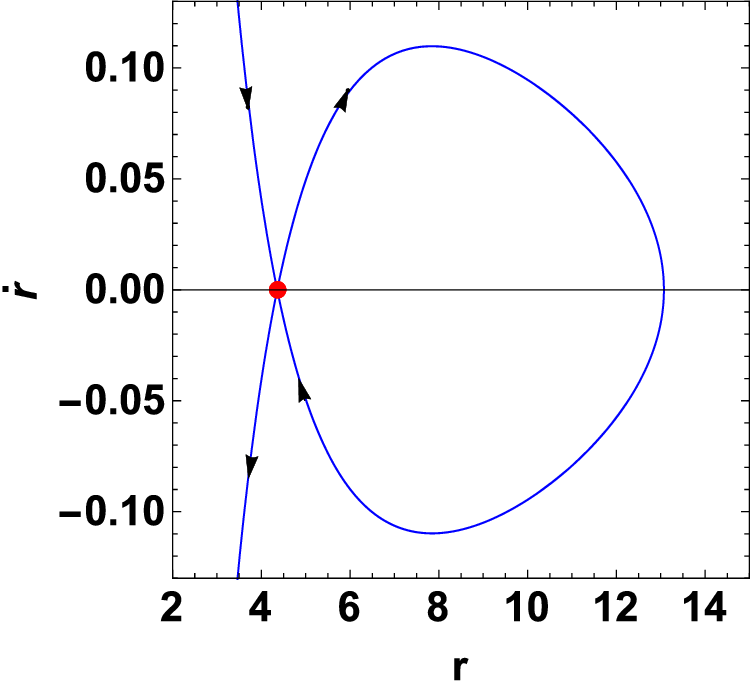}\;\;\;\;\;\includegraphics[width=5.5cm ]{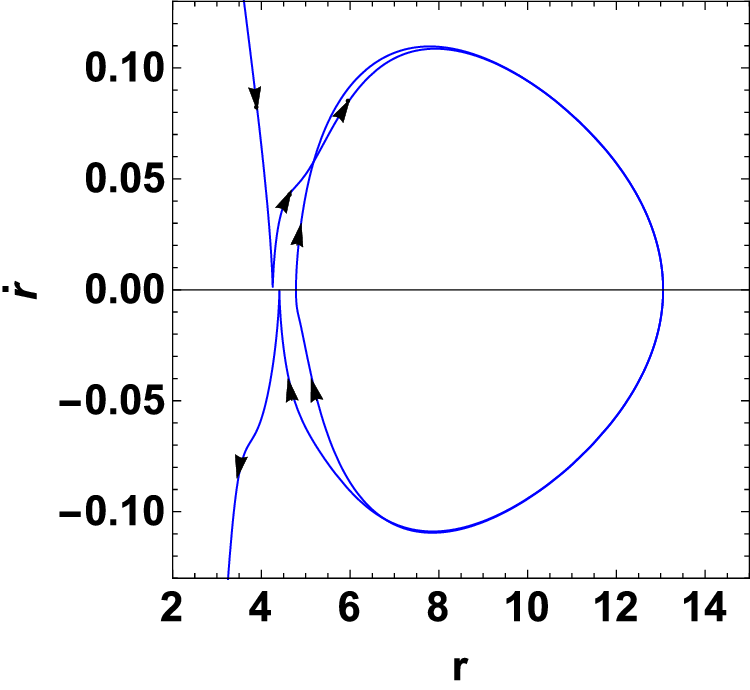}
\caption{The hyperbolic fixed point (red point) in Kerr spacetime ( in the left panel) in ref.\cite{Rana} and the homoclinic tangles under the interaction with Chern-Simons invariant with $\alpha=4$ ( in the right panel).}\label{fign12}
\end{figure}
In Fig.\ref{fign12}, the red point is the hyperbolic fixed point. The stable manifold and the unstable manifold along homoclinic orbit H3 are plotted in the left panel for the case without the coupling, while in the right panel, we present the stable manifold and the unstable manifold under the interaction with the coupling constant $\alpha=4$. It is shown that there exists intersection between these two manifolds as $\alpha=4$, which means that homoclinic tangles occurs and then there always exists chaos in the dynamical system (\ref{Hamin}) with the $\alpha\neq0$.

Let us now to study chaotic motion of scalar particles coupling to Chern-Simons invariant in Kerr black hole spacetime through the Poincar\'e section, the fast Lyapunov indicator, bifurcation diagram  and basins of attraction.
In order to probe motion of scalar particle coupling to Chern-Simons invariant in Kerr black hole spacetime, we must solve numerically differential equations (\ref{wfE1})-(\ref{wfE3}). Here, we adopt to the corrected fifth-order Runge-Kutta method \cite{Huang,DZMa1,DZMa2,DZMa3}, where the high precision can be effectively ensured by correcting the velocities $(\dot{r}, \dot{\theta})$ at every integration step so that the numerical deviation is pulled back in a least-squares shortest path.
For a dynamical system,  the motion orbit of the particle is entirely determined by its initial conditions and the parameters of system. In principle, the choice for the parameters and initial conditions of particle should be arbitrary.
For a convenience, we here set the initial motion orbit of the particle as a periodic orbit in the zero-coupling case and then
probe the change of the degree of disorder of particle orbits with the coupling parameter $\alpha$.
\begin{figure}
\includegraphics[width=5cm ]{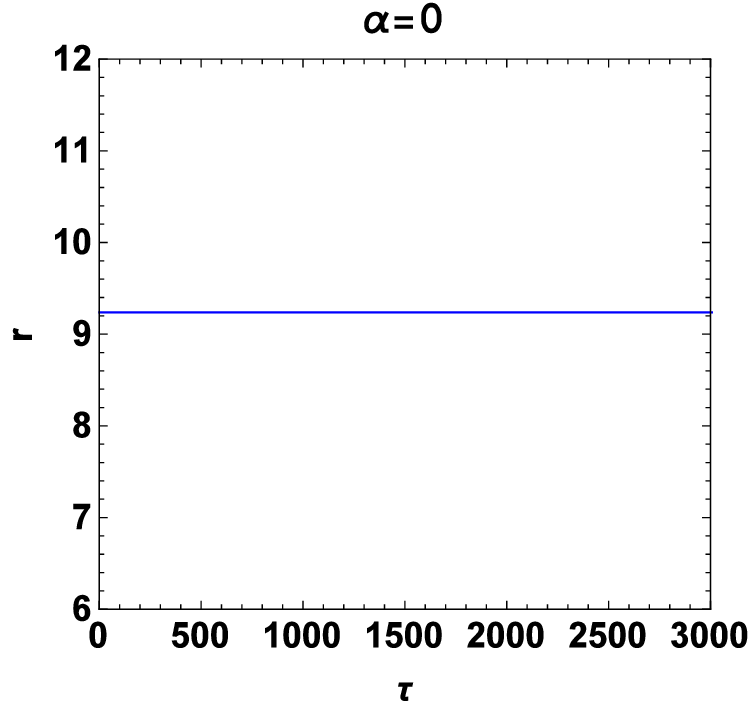}\includegraphics[width=5cm ]{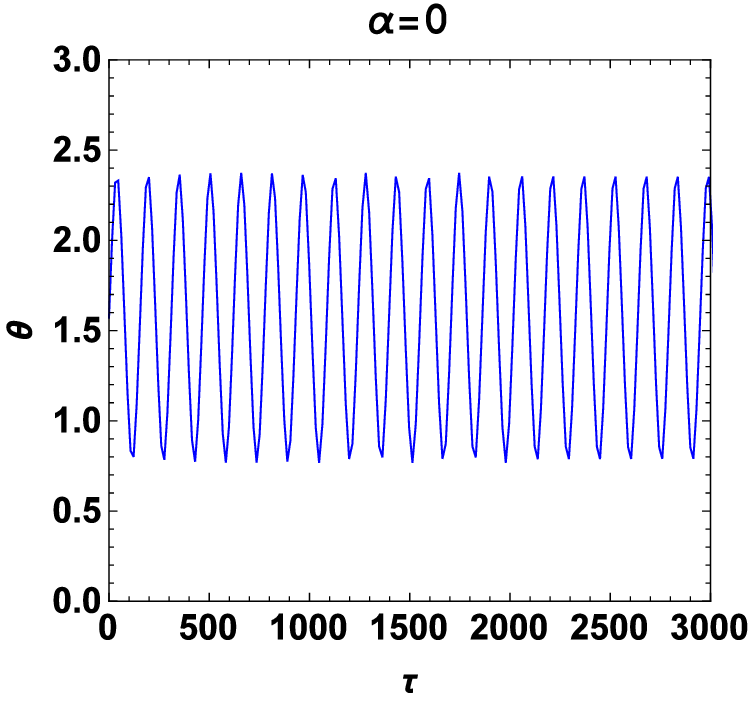}\;\;\;\;\;\includegraphics[width=5cm ]{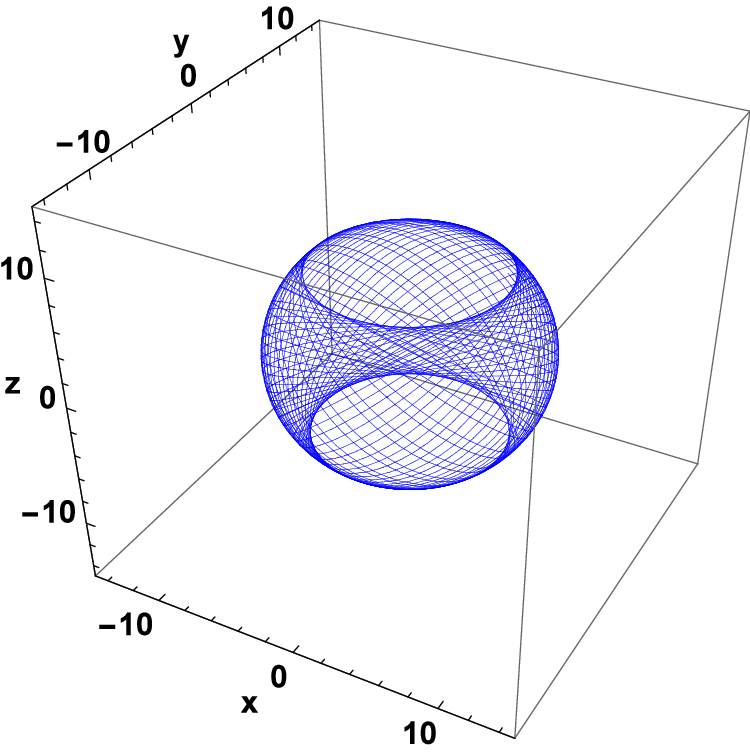}
\caption{The chosen periodic orbit of the scalar particle in a Kerr black hole spacetime with $E=0.95$, $L=2.4M$, $a=0.8$ and $\alpha=0$.}\label{fig1}
\end{figure}
The chosen periodic orbit for the scalar particle is shown in Fig.\ref{fig1},  which can be obtained with the parameters
$\{$$E=0.95$, $L=2.4M$, $a=0.8$, $\alpha=0$$\}$ and the initial conditions
$\{$$r(0)=9.238142$, $\dot{r}(0)=0$, $\theta(0) =\frac{\pi}{2}$$\}$. For this orbit, the radial coordinate $r(\tau)$ is fixed, but the polar angle $\theta(\tau)$ changes periodically with the affine parameter $\tau$. Thus, it belongs to a family of non-planar spherical orbits as shown in the right panel in Fig.\ref{fig1}.

We are now in position to study the effect of the coupling with Chern-Simons invariant on this  periodic orbit of scalar particle in Kerr black hole spacetime.
\begin{figure}[ht]
\includegraphics[width=16cm ]{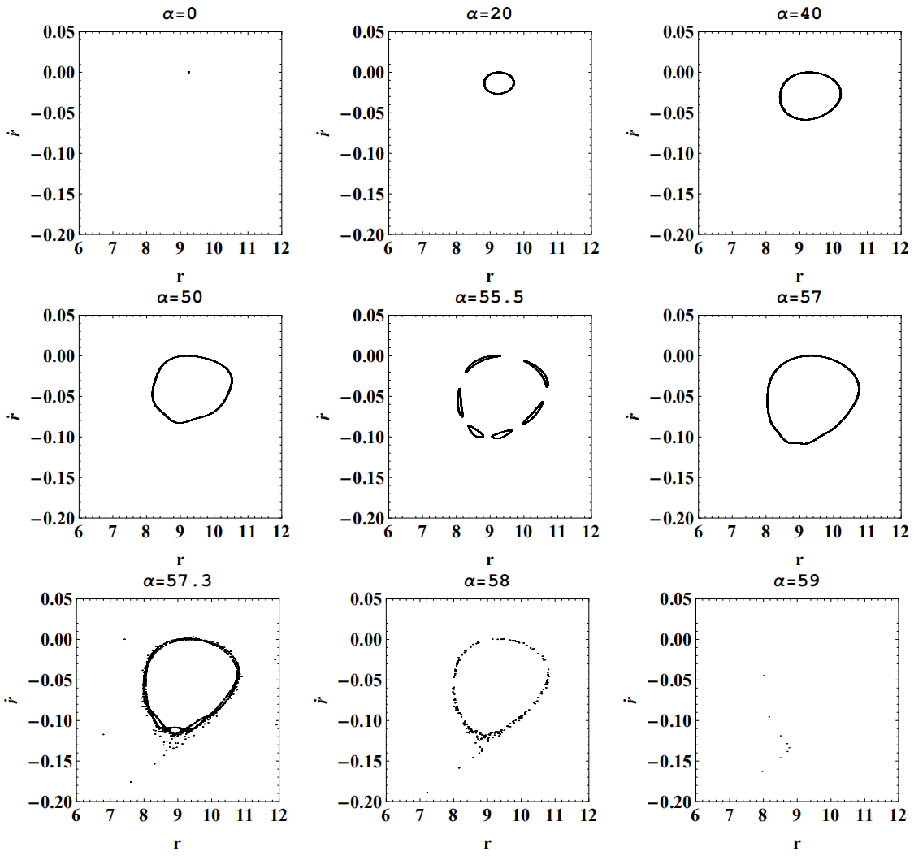}
\caption{The Poincar\'e surface of section $(\theta=\frac{\pi}{2})$ in the plane $r-\dot{r}$ with different $\alpha$ for the signal plotted in Fig.\ref{fig1}. }\label{fig5}
\end{figure}
\begin{figure}[ht]
\includegraphics[width=16cm ]{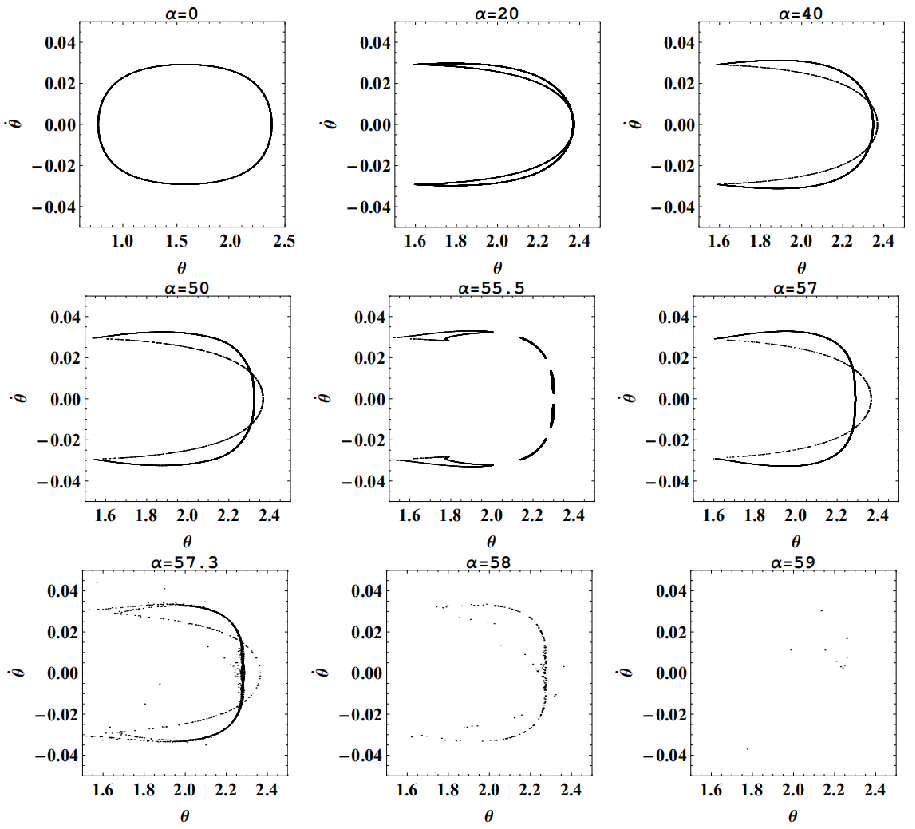}
\caption{The Poincar\'e surface of section $(\dot{r}(\tau)=0$) in the plane $\theta-\dot{\theta}$ with different $\alpha$ for the signal plotted in Fig.\ref{fig1}. }\label{figs501}
\end{figure}
It is well known that Poincar\'e section is an effective method to discern chaotic motion. It is an intersection of particle's trajectory and a given hypersurface which is transversal to the trajectory in the phase space. According to the intersection point distribution in Poincar\'e section, the motions of particles  are classified as three kinds for a dynamical system.
The periodic motions and the quasi-periodic motions correspond to a finite number of points  and a series of close curves in the Poincar\'e section, respectively.  The chaotic motion solutions correspond to strange patterns of dispersed points with complex boundaries \cite{pjl}. In Fig.\ref{fig5}, we show the change of the Poincar\'e sections $(\theta=\frac{\pi}{2})$ in the plane $r-\dot{r}$ with different Chern-Simons coupling parameter $\alpha$. As $\alpha=0$, we find that there is only a point in the Poincar\'e sections $(\theta=\frac{\pi}{2})$, which confirms again that the orbit in Fig.\ref{fig1} is periodic.
For the cases with $\alpha\neq0$, it is shown that the phase path of the coupled scalar particle differs from that in the case without coupling. From Fig.\ref{fig5}, one can find that for $\alpha<57.3$ the phase path of the coupled scalar particle is a quasi-periodic  Kolmogorov-Arnold-Moser (KAM) tori and the corresponding motion is regular under the interaction with the Chern-Simons invariant. Moreover, we find that the KAM tori deforms with the increase of  $\alpha$.
Especially, as $\alpha=55.5$, there is a chain of islands which is composed of six secondary KAM toris belonging to the same trajectory. With the coupling constant $\alpha$ increasing, the chain of islands are joined together and become a big KAM tori. This means that trajectory of the scalar particle coupled to the Chern-Simons invariant is regular  as $\alpha<57.3$ since a regular orbit moves on a torus in the phase space and the corresponding curve in Poincar\'e sections is a cross section of the torus. However, as the Chern-Simons coupling parameter increases further to $\alpha=57.3$, the KAM tori is broken and many discrete points distribute randomly in the section. Combining with the Fast Lyapunov indicators (FLI) shown in Fig.\ref{fig4}, one can find that the FLI($\tau$) grows with exponential rate in the case with $\alpha=57.3$ and then the motion is chaotic since
the exponential growth of FLI($\tau$) means that the motion of the coupled particle is very sensitive to initial value, which is a feature of chaotic motion. The presence of chaos implies that the
motion of the coupled particle (\ref{Hamin}) is non-integrable because chaos appears only in the non-integrable system in which the number of independent constants of the motion  is less than degrees of freedom.
As $\alpha=59$, one can find that the tori is completely vanished and there exist a few discrete
points distributed randomly in Poincar\'e section. It is caused by that the particle falls finally into the event horizon of the black hole or spatial infinite after undergoing some chaotic oscillations around black hole, which is different essentially from those in the case of usual multiple-periodic motion. Thus, the scalar particle becomes more complicated under the coupling with the Chern-Simons invariant. In Fig.\ref{figs501}, we also present the Poincar\'e section in the plane $\theta-\dot{\theta}$ with the condition $(\dot{r}(\tau)=0$). As $\alpha=0$, one can find that the phase path is a closed curve rather than only a point for the periodic signal plotted in Fig.\ref{fig1}. It is not surprising because for the particle moving along this periodic orbit, its polar coordinate $r$ is a constant and the condition $\dot{r(\tau)}=0$ is always satisfied, which leads to that the phase path in the Poincar\'e section $(\dot{r}(\tau)=0$) is a closed curve in the plane $\theta-\dot{\theta}$. In the non-zero $\alpha$ cases, we find that the chain of islands appears as $\alpha=55.5$ and the KAM tori is destroyed as $\alpha=57.3$, which are consistent with those in the Poincar\'e section $r-\dot{r}$.  Thus, the properties of orbits of particle's motion reflected by the phase paths are same in both kinds of Poincar\'e sections.
\begin{figure}[ht]
\includegraphics[width=16cm ]{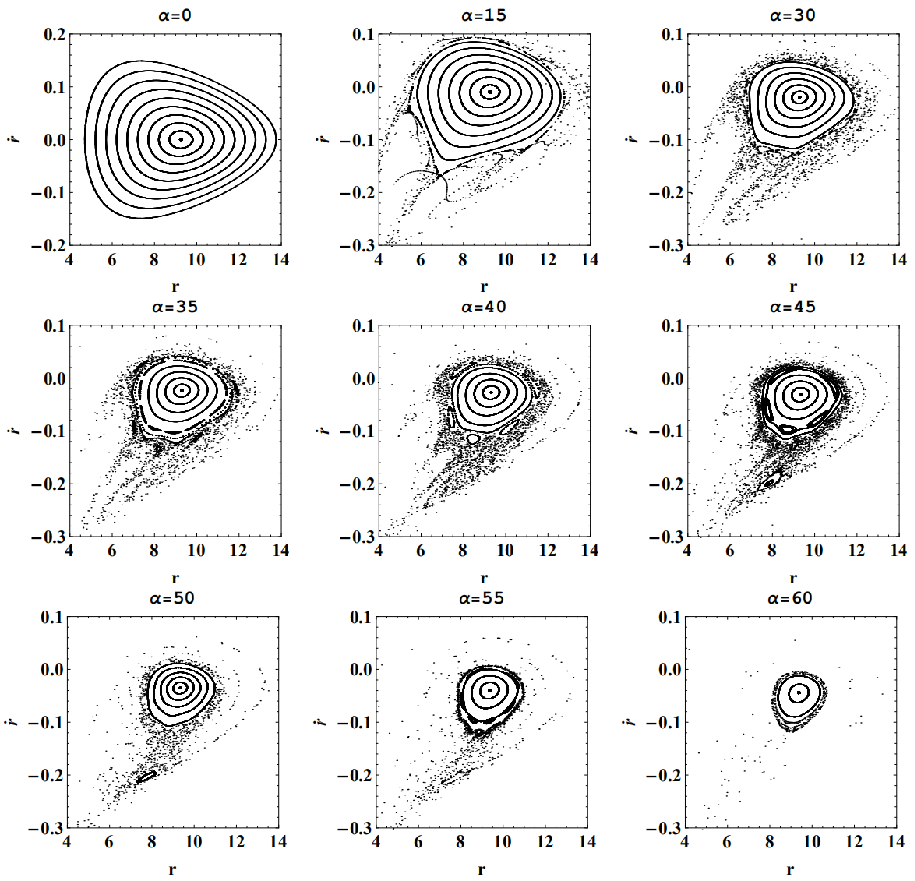}
\caption{The change of Poincar\'e section ($\theta=\frac{\pi}{2}$) with the coupling parameter $\alpha$ for the motion of the scalar particle coupled to Chern-Simons invariant in a Kerr black hole spacetime with the fixed  $ a=0.8$, $ E=0.95$ and $L= 2.4M$. }\label{fig6}
\end{figure}

\begin{figure}[ht]
\includegraphics[width=16cm ]{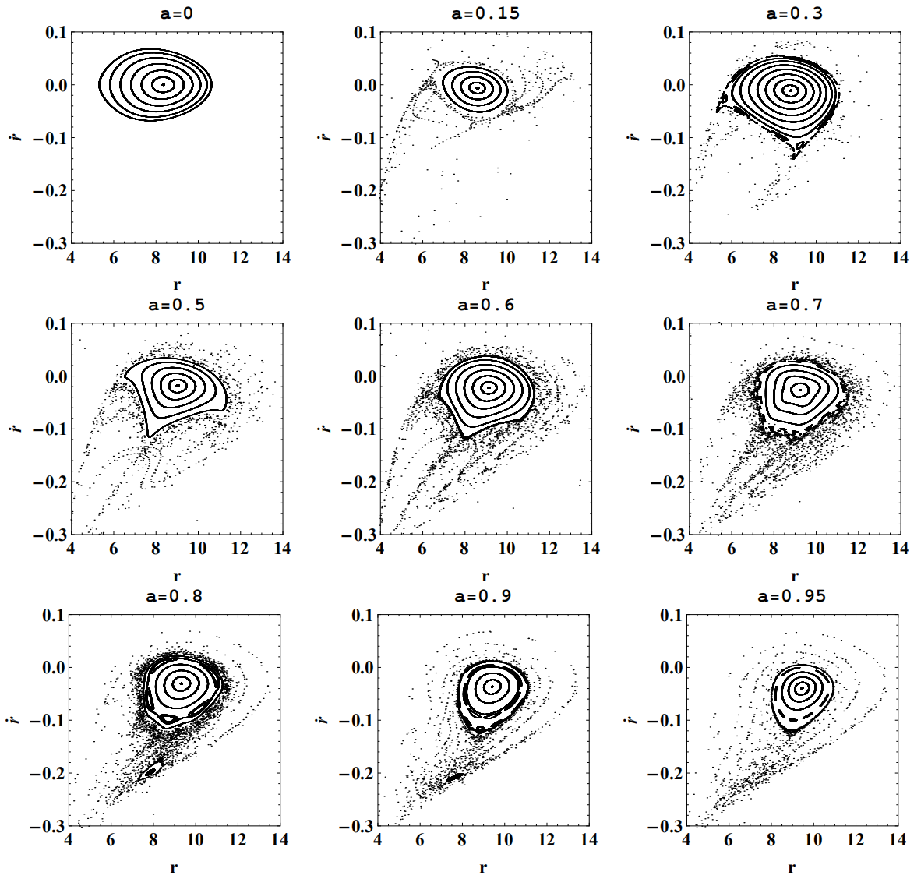}
\caption{The change of Poincar\'{e} section ($\theta=\frac{\pi}{2}$) with the spin parameter $a$ for the motion of the coupled scalar particle in a Kerr black hole spacetime with the fixed parameters $\alpha=45$, $E=0.95$ and $L=2.4M$.}\label{fig7}
\end{figure}

\begin{figure}
\includegraphics[width=16cm ]{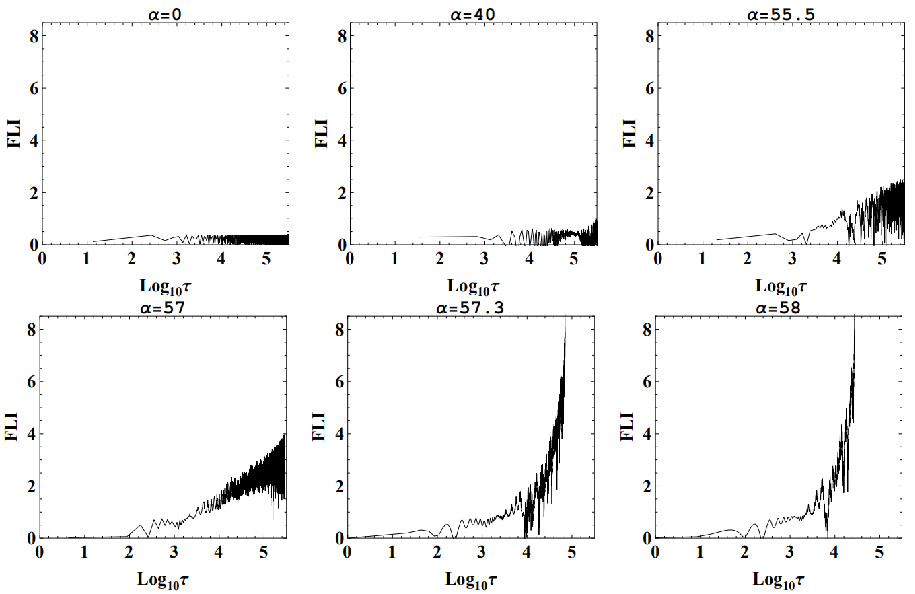}
\caption{The Fast Lyapunov indicators  with different $\alpha$ for the signal plotted in Fig.\ref{fig1}.  }\label{fig4}
\end{figure}

In Figs.(\ref{fig6}) and (\ref{fig7}), we present Poincar\'e section $(r, \dot{r})$ containing more motion orbits of
the coupled particle in the background of Kerr black hole spacetime. We find that for the fixed spin parameter $a=0.8$, with the Chern-Simons coupling parameter $\alpha$, the main island of stability shrinks , while the chaotic orbits are driven faster towards the horizon. We also note that in the case with $\alpha=0$, there are a series of close curves in the Poincar\'e section and then there exists only regular orbits for the scalar particle in the Kerr black hole spacetime. The main reason is that in the case without coupling the motion equations of scalar test particles reduce to the usual variable-separable geodesic equations. For the fixed coupling parameter $\alpha=45$, we find that the chaotic region increases with the increase of the spin parameter $a$. Especially, as $a=0$, all of the orbits of particles is regular, which can be explained by a fact that the Chern-Simons invariant $^* R R$ disappears and the motion of particle reduces to that of in the usual Schwarzschild black hole spacetime. These result indicate clearly in the Poincar\'e section that the numbers and
positions of fixed points of the system depend on the spin parameter $a$ and the Chern-Simons coupling parameter $\alpha$.
Moreover, from Figs. \ref{fig5}-\ref{fig7}, we note that the patterns in the Poincar\'e section own the reflection symmetry along line $\dot{r}=0$ in the zero-coupling case. However, this symmetry is lost with the presence of the coupling. It is not surprising because the Chern-Simons invariant $^* R R$ (\ref{CSRR}) is not symmetric with respect to the equatorial plane. Under the interaction with the Chern-Simons invariant $^* R R$, the orbits of particles lose the symmetry with respect to the equatorial plane and then the radial velocity distribution of particle crossing the equatorial plane has not the reflection symmetry along the line $\dot{r}=0$.

FLI is a kind of fast and effective tools to discern chaotic behavior of particle \cite{Froesche,FroeschE,Fouchard,Tancredi}. In the curved spacetime, the FLI with two particles method can be described by \cite{Wu5,Wu51,Wu52}
\begin{eqnarray}
FLI(\tau)=-k[1+\log_{10}d(0)]+\log_{10}\bigg|\frac{d(\tau)}{d(0)}\bigg|,
\end{eqnarray}
where $d(\tau)=\sqrt{|g_{\mu\nu}\Delta x^{\mu}\Delta x^{\nu}}|$, $\Delta x^{\mu}$ denotes the deviation vector between two nearby
trajectories at proper time $\tau$.  The sequential number of renormalization $k$ is used to avoid the numerical saturation arising from the fast separation of the two adjacent orbits. It is well known that FLI$(\tau)$ grows with exponential rate for chaotic motion, even for weak chaotic motion, but it grows algebraically with time for the
regular resonant orbit and for the periodic one. In Fig.\ref{fig4}, we present the change of FLI$(\tau)$ with the coupling parameter $\alpha$ for the chosen initial orbit of scalar particle plotted in Fig.\ref{fig1}, which tells us that in the cases with $\alpha<57.3$, the FLI($\tau$) increases linearly with $\tau$ and then the  motions of the scalar particle are regular. However, in the case with $\alpha=57.3 $ or $58$, the FLI($\tau$) grows with exponential rate and then the corresponding motions are chaotic. These results are consistent with those obtained by the Poincar\'e section shown in Fig.\ref{fig5}.

The bifurcation diagram can show how the motion orbit of the scalar particle depend on the spin parameter $a$ of black hole  and the Chern-Simons coupling parameter $\alpha$. Here, we present the bifurcation of the radial motion for the coupled scalar particle.  \begin{figure}[ht]
\includegraphics[width=16cm ]{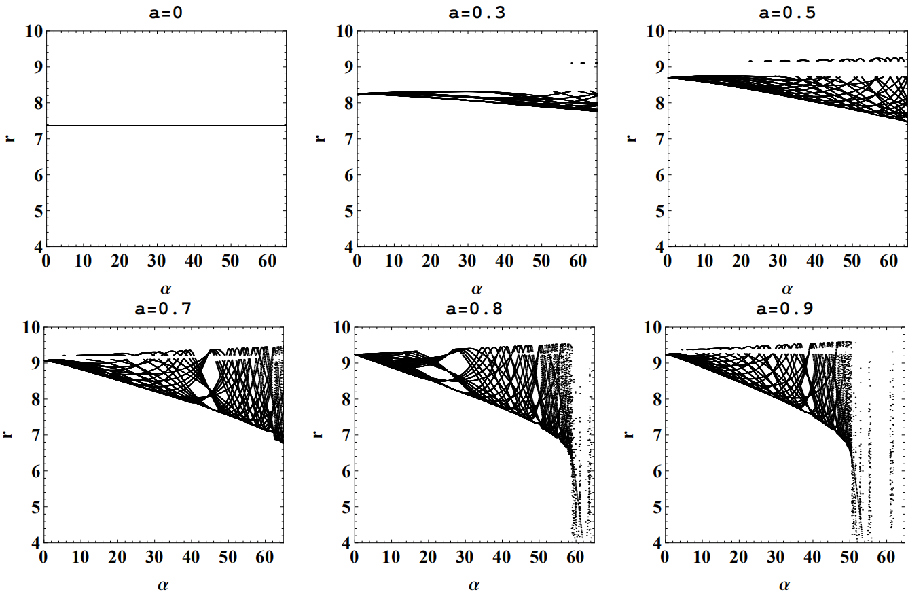}
\caption{The change of bifurcation with the Chern-Simons coupling parameter $\alpha$
for the signal in Fig.\ref{fig1}. }\label{fig8}
\end{figure}
\begin{figure}[ht]
\includegraphics[width=16cm ]{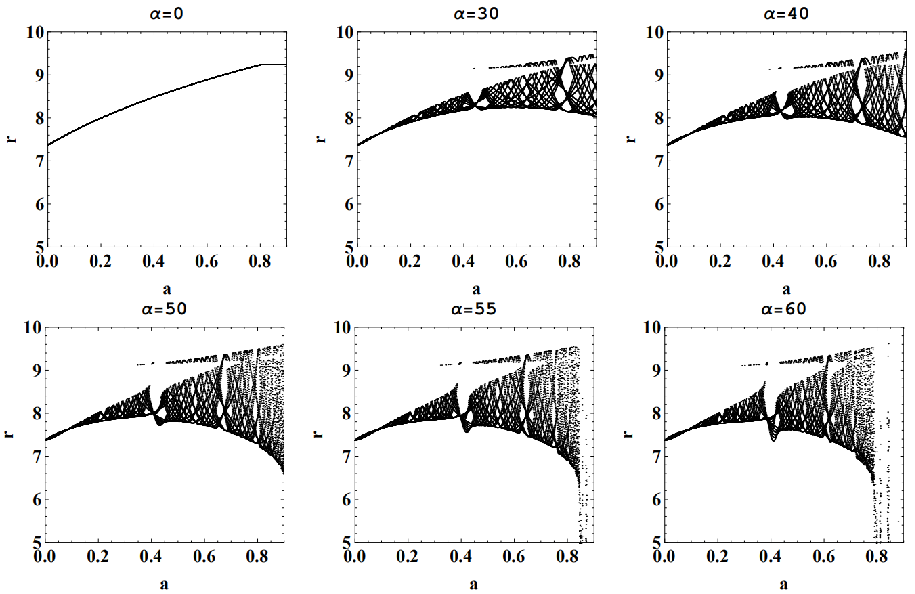}
\caption{The change of bifurcation  with the spin parameter $a$ for the signal in Fig.\ref{fig1}.}\label{fig9}
\end{figure}
In Figs. \ref{fig8} and \ref{fig9}, we plot the
bifurcation diagram of radial coordinate $r(\tau)$ with the Chern-Simons coupling parameter $\alpha$ and the
spin parameter $a$ for the scalar particle motion the signal in Fig.\ref{fig1} in a Kerr black hole
spacetime. Here we chose the initial orbit plotted in Fig.\ref{fig1}. As $a=0$ or $\alpha=0$,  one can find that the radial coordinate $r(\tau)$ is a periodic function and there is no bifurcation for the dynamical system, which means that the motions of the scalar particles are regular in these two cases. The non-existence of bifurcation for the scalar particle in the Schwarzschild case is caused by a fact that the Chern-Simons invariant $^* R R$ (\ref{CSRR}) vanishes as the spin parameter $a=0$. In the rotating black hole case, we find that the radial motions of particles is dominated by
the Chern-Simons coupling parameter $\alpha$ and the spin parameter $a$.  From Fig.(\ref{fig8}),
for the chosen initial signal plotted Fig.\ref{fig1}, one can find that with the increase of the spin parameter $a$ the range of $\alpha$ chaotic motion appeared increases. Fig.\ref{fig9} shows that the lower bound $a$ of emerging chaotic orbits decreases with the Chern-Simons coupling parameter $\alpha$.  These indicate that the effects of the Chern-Simons coupling parameter and the spin parameter on the motion of the coupled scalar particle are very complex, which are typical features of bifurcation diagram for the usual chaotic dynamical system. Thus, the dynamical behaviors of scalar particle, under the interaction with Chern-Simons invariant, becomes much richer in a Kerr black hole spacetime.

The fractal basin boundary can provide an independent signature of chaos \cite{Cornish,Frolov,Zayas,Basin,Basin2}. It is very common for a dynamical system to own more than one attractor. For each such attractor, its basin of attraction is the set of initial conditions in the phase space whose trajectories approaches that attractor. The boundary between basins can be either smooth or fractal, which depends on dynamical properties of system. Actually, the fractal of basin boundary is a result of chaotic motion of orbits on the boundary. Thus, one can identify whether chaos is present or not by analysing the basin boundary of attraction.
\begin{figure}[ht]
\includegraphics[width=3.95cm ]{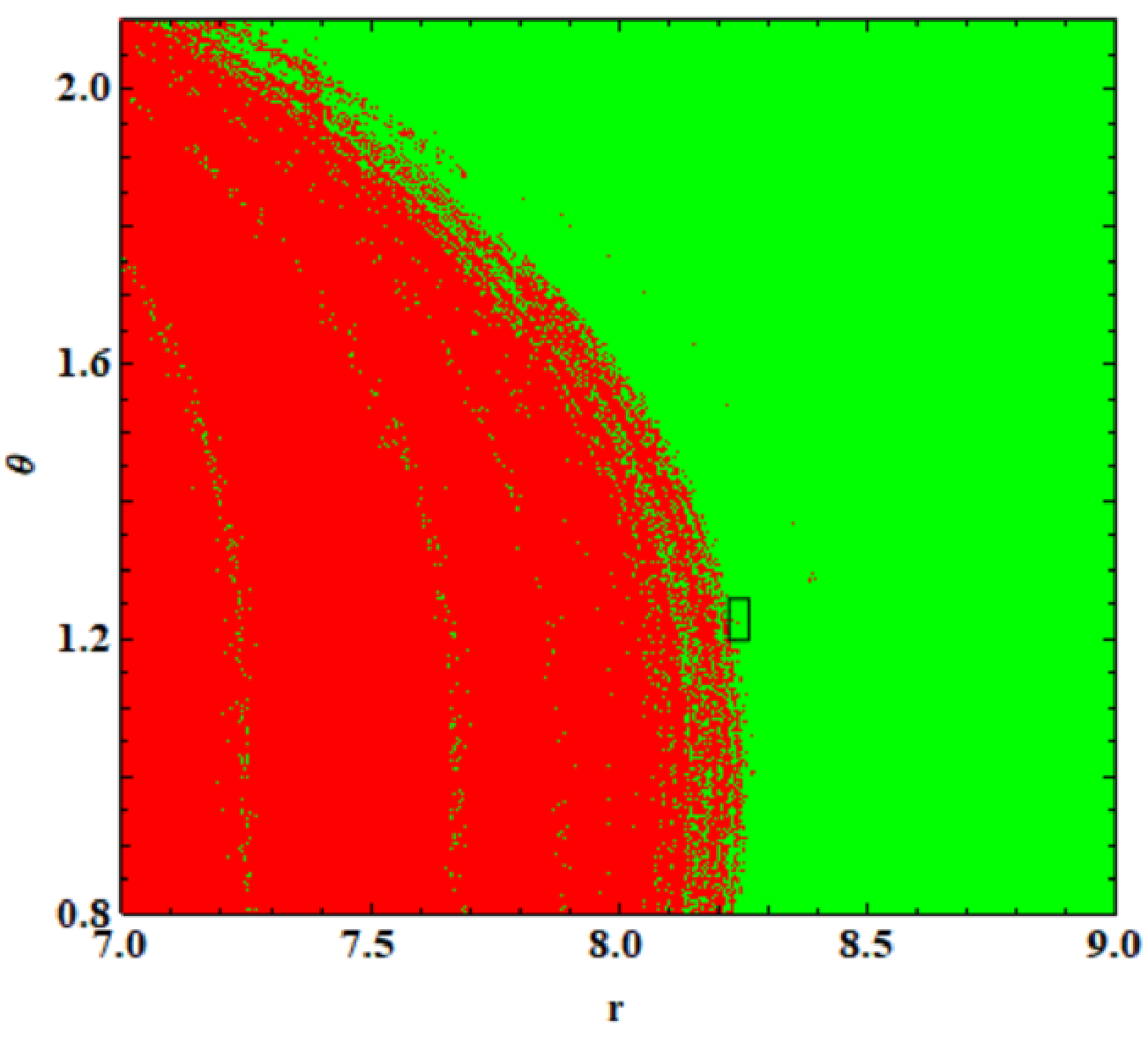}
\includegraphics[width=4.1cm ]{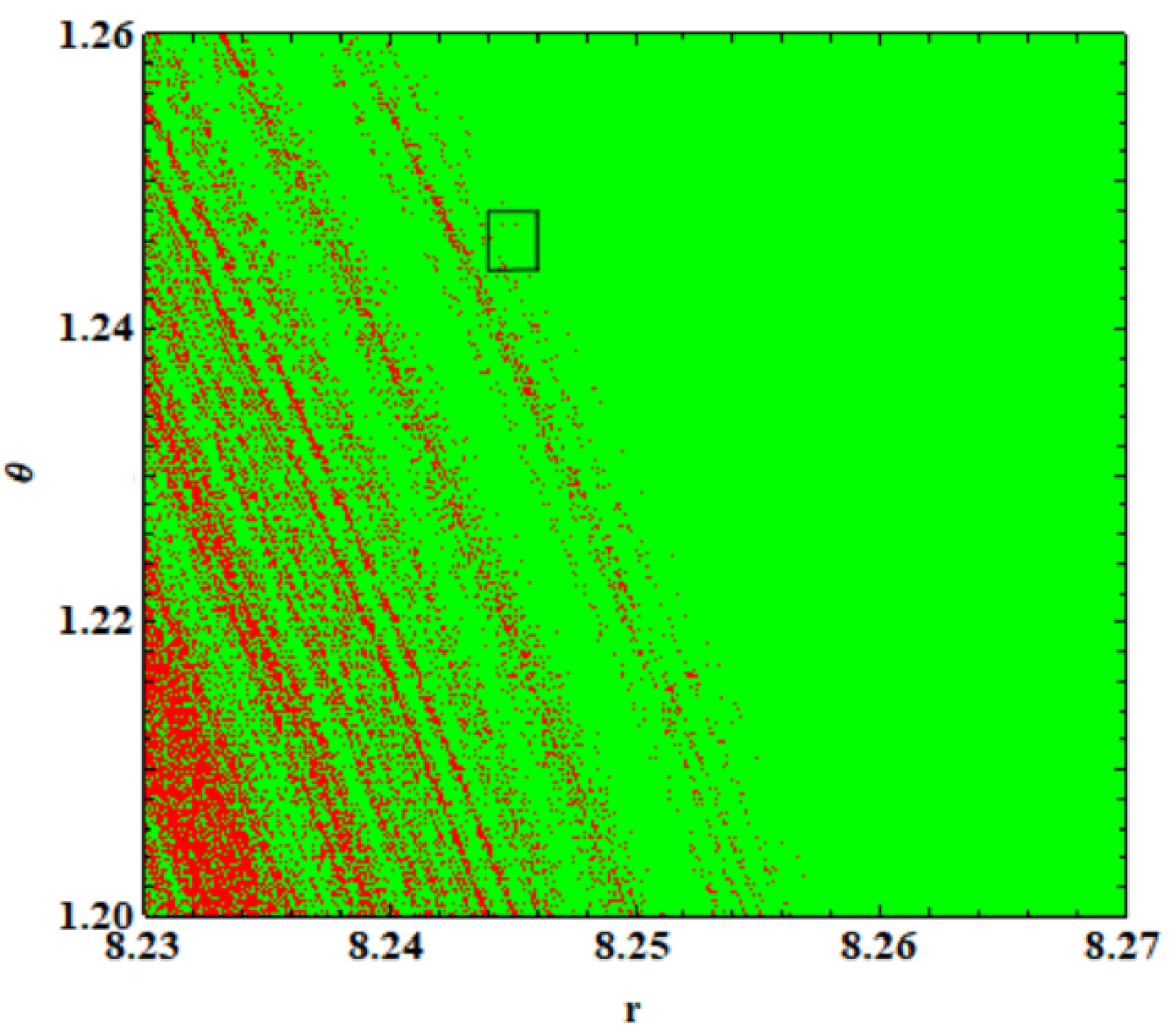}
\includegraphics[width=4.15cm ]{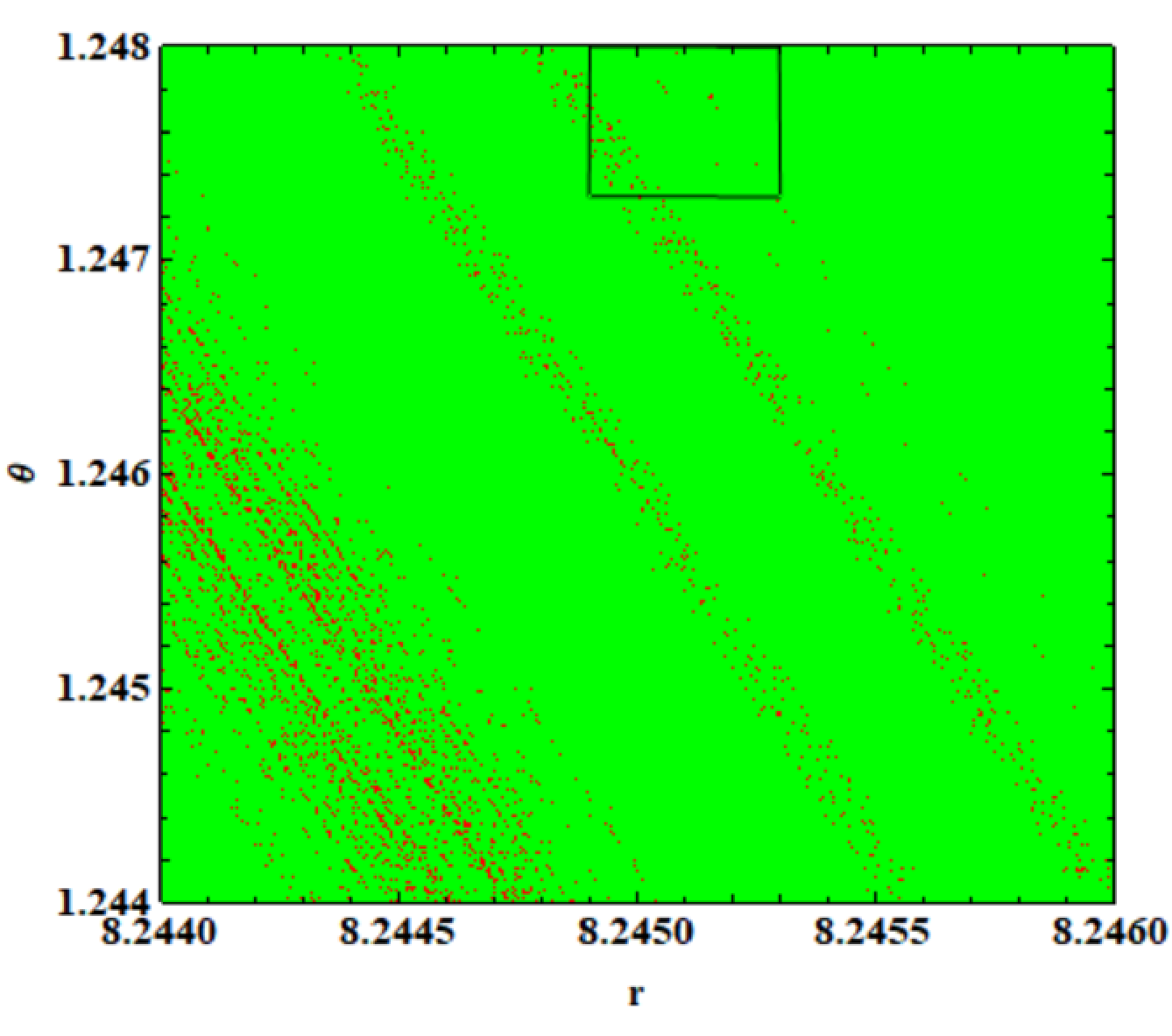}
\includegraphics[width=4.22cm ]{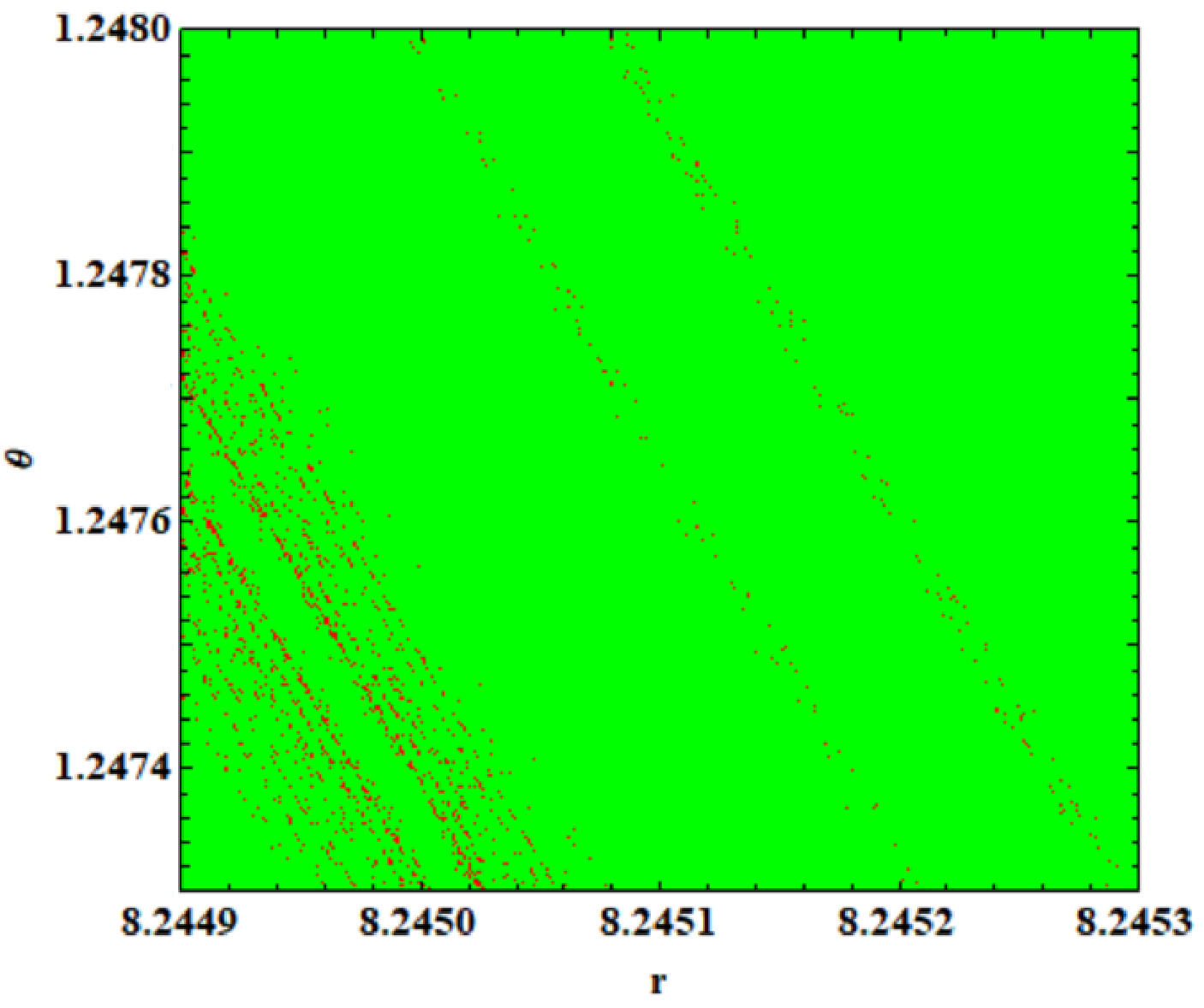}
\caption{The fractal basins of attraction for the coupled scalar particle in a Kerr black hole spacetime with the fixed parameters $a=0.8$, $\alpha=45$, $E=0.95$ and $L=2.4M$. }\label{fig11}
\end{figure}
In Fig. \ref{fig11}, we plot the basins of attraction in a large subset of phase space for the coupled scalar particle in a Kerr black hole spacetime with the fixed parameters $a=0.8$, $\alpha=45$, $E=0.95$ and $L=2.4M$. All these
initial conditions correspond to the values of $r$ and $\theta$ shown in the figure, $p_r=0$ and $p_{\theta}$ given
by the constrain (\ref{Hcon}), i.e., $h=0$. The red point corresponds to the case where the geodesic crosses the event horizon and the particle falls into black hole. The blue point denotes that the geodesic
radially escapes to infinity, while green point corresponds to that the geodesics settles down to an oscillatory motion of particle. Here, the condition for capture is simply $r\leq r_H$ and the condition for escape is set as $r\geq100r_H$. For green points, we consider trajectories that neither were captured nor escaped to infinity after 50000 iterations. From Fig. \ref{fig11}, it is easy to obtain the basins boundaries possess a self-similar fractal fine structure, which also means that there exists chaotic motion for a coupled scalar particle in Kerr black hole spacetime under the interaction with the Chern-Simons invariant.

\section{Summary}

In this paper we present firstly  the equation of motion for
a test scalar particle coupled to the Chern-Simons invariant in the Kerr black hole spacetime through the
short-wave approximation. We have analyzed the dynamical
behaviors of the test coupled particles
by applying techniques including Poincar\'e sections,  fast Lyapunov exponent indicator, bifurcation
diagram and basins of attraction.  Our results confirm that there exists chaotic phenomenon in the motion of scalar particle interacted
with the Chern-Simons invariant in the background of a rotating black hole. The main reason is that such coupling leads to that the equation of motion is not variable-separable and the dynamical system
is non-integrable. Moreover, we probe effects of the coupling parameter and black hole spin parameter
on the motion of test coupled scalar particle.
For the fixed spin parameter $a=0.8$, we find that the main island of
stability in the Poincar\'e section shrinks with the Chern-Simons coupling parameter $\alpha$, while the chaotic orbits are driven faster towards  the horizon. For the fixed coupling parameter $\alpha=45$, we find that the chaotic region increases  with the black hole spin parameter $a$.  Moreover, as $a=0$, all of the orbits of particles is regular, which can be explained by a fact that the Chern-Simons invariant $^* R R$ disappears and the motion of particle reduces to that of in the case without the coupling. These mean that the coupling between scalar particle and Chern-Simons invariant yields the richer dynamical behavior of
scalar particle in a Kerr black hole spacetime.

\section{\bf Acknowledgments}

This work was  supported by the National Natural Science
Foundation of China under Grant No.11875026, 11875025, 12035005 and 2020YFC2201403.

\end{document}